# Experimental setups for XPS measurements beyond the instrumental lateral resolution limit

Uwe Scheithauer, ptB, 82008 Unterhaching, Germany
E-Mail: scht.ptb@t-online.de; scht.uhg@mail.de
Internet: orcid.org/0000-0002-4776-0678; www.researchgate.net/profile/Uwe_Scheithauer

## Abstract:

The lateral resolution of an X-ray photoelectron spectroscopy instrument, which is equipped with a focused X-ray beam, is limited by the nominal X-ray beam diameter and the long tail intensity distribution of the X-ray beam. The long tail intensity distribution of the X-ray beam impedes to perform a measurement with good lateral resolution and low detection limits at the same time.

Two experimental setups are described which allow examining sample structures that are smaller than the X-ray beam dimensions. The first method uses differential sample charging on partly non-conductive samples by low energy electron flooding. The spectra of the non-conductive sample areas are shifted towards lower binding energy. That way, the surface compositions of conductive and non-conductive sample areas are estimated independently. The second method utilizes the rather limited dimensions of the energy analyser acceptance volume. Here only the sample is placed inside the energy analyser acceptance volume. That way, signals from the illuminated sample contribute exclusively to the measured photoelectrons intensity, independent form the sample size.

## 1. Introduction

The resolvable sample structures of X-ray photoelectron spectroscopy (XPS) microprobes, which are equipped with a focused X-ray beam, are defined by the minimal X-ray beam diameter of this focused X-ray beam. This minimal X-ray beam diameter is given by the beam width at the half maximum intensity level or a similar vendor's definition. Such a definition of the beam diameter ignores the long tail intensity distribution of the X-ray beam [1]. In case of trace element detection on small sample features it is impossible to decide whether this signal refers to a contamination of the small sample feature or if it comes from the surrounding of the small sample feature due to the long tail intensity. Therefore an XPS measurement with higher lateral resolution using an instrument with a focused X-ray beam is limited by the nominal X-ray beam diameter and the long tail intensity distribution of the X-ray beam.

In this article, two approaches are presented, which overcome this limitation by dedicated experimental setups. The approaches allow examining sample structures, which are smaller than the X-ray beam dimensions. The first method utilizes differential sample charging on partly non-conductive samples to discriminate conductive and non-conductive sample regions [2]. The second method utilizes the limited dimensions of the energy analyser acceptance volume. If exclusively the sample is placed inside this volume, this way only the illuminated sample contributes to the measured photoelectron intensity, independent form the X-ray beam diameter and the sample size. For each approach a measurement example is shown.

## 2. Instrumentation

Two XPS microprobes were used for the XPS measurements presented here: A Surface Science Laboratories X-probe instrument and a Physical Electronics XPS Quantum 2000, respectively. Both XPS instruments achieve the spatial resolution by the combination of a fine-focused electron beam generating the X-rays on a water cooled Al anode and a mirror quartz monochromator, which monochromatizes the Al$_{k\alpha}$ radiation and refocuses the X-rays to the sample surface. This way the beam diameter of the electron beam, which generates the X-rays on the Al anode, defines the X-ray beam diameter on the samples surface. Both instruments have a low voltage electron gun for charge compensation and a differentially pumped Ar$^+$ ion gun for sample cleaning and sputter depth profiling.

The Surface Science Laboratories instrument is equipped with a 6'' wafer sample handling system. Fig. 1 shows a sketch of the instrument. Samples features are selected and adjusted at the instruments measurement position utilizing an optical microscope. The instrument has nominal X-ray spot sizes of 150, 300, 600 and 1000 μm. The X-ray beam, the mean take-off angle of the energy analyzer and the Ar$^+$ ion gun have an angle of 55° relative to the surface normal of a flat mounted sample. The hemispherical energy analyzer has a collection lens with 30° acceptance angle and multichannel detection via a micro-channel plate electron multiplier and a position sensitive detector. A more detailed description of the instrument is given in literature [3].

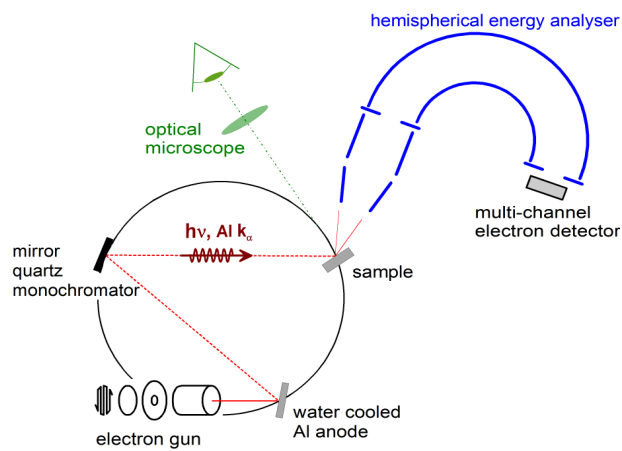

Fig. 1: Sketch of Surface Science Laboratories XPS instrument

The Physical Electronics XPS microprobe Quantum 2000 represents the further development of the primary focused XPS microprobe concept. By controlling the electron beam diameter, nominal X-ray beam diameters between 10 μm and 200 μm are selectable with the instrument used here. The X-ray beam scans across the sample as the electron beam is scanned across the Al anode by applying electrostatic deflection voltages to the electron beam. On the sample surface at maximum an area of approximately 1.4 x 1.4 mm$^2$ can be scanned (see fig 2). With a fine focused rastered X-ray beam a sample can be depicted by an X-ray beam induced secondary electron image. This way, sample features are localized. An optical microscope utilized for sample inspection at the measurement position becomes unnecessary thanks to this imaging capability. Therefore, in a Quantum 2000, the space directly above the sample is used for the X-ray source mounting. For a flat mounted sample as used here in a Quantum 2000 the incoming X-rays are parallel to the surface normal. In this geometrical situation, the mean energy analyser take-off axis and the $Ar^+$ ion gun are oriented approximately 45° relative to the sample surface normal. The hemispherical energy analyzer has a collection lens with an acceptance angle of 30° and a 16 channel detector. A more detailed description of the instrument and its performance is given in literature [1, 4-10].

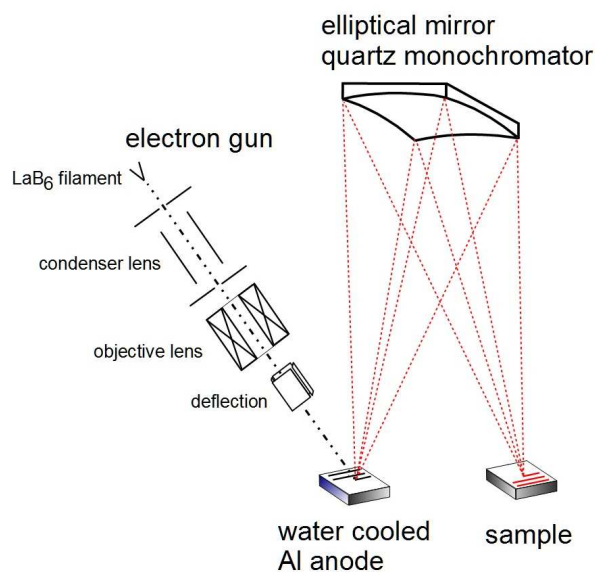

Fig. 2: imagining X-ray source of a Quantum 2000

## 3. Quantitative Lateral Resolution of an X-ray Microprobe

The X-ray beam diameters of XPS microprobes with a focused X-ray beam are measured using a vendor-specified measurement procedure. For the Quantum 2000 the manufacturer defines this beam size as the distance between the points at which the signal amplitude is 20% and 80% of the maximum value when the beam is scanned over a material edge [6]. However, for a more precise determination of the intensity distribution within the X-ray beam the long tail contributions of the X-ray beam have to be taken into account [1].

The long tail intensity was measured using Pt apertures of different diameters, which were mounted over drilled holes (see insert of fig. 3). If the X-ray beam is centred in the aperture, only long tail X-ray beam intensities from outside the aperture diameter produce a Pt signal. This measured Pt intensity is normalised by a second measurement on massive Pt. The use of apertures as test samples has several advantages. Details are discussed in literature [1].

Fig. 3 shows the results of these measurements for X-ray beams with 4 different nominal X-ray beam diameters. For aperture diameters, which are only 2...4 times larger than the beam size, the normalized signal drops down as expected. For instance for a small beam with 16 µm diameter as estimated by the 20% to 80% edge test method as much as 10% of the signal comes from outside a 50 µm aperture. For apertures having a diameter of 300 µm or more the normalized signal is independent from the beam diameter and decreases slowly. Per definition the quantitative lateral resolution of an X-ray microprobe is the diameter of an aperture which produces a normalised intensity of 1 %. It is approximately 450 µm for the Quantum 2000 used here [1].

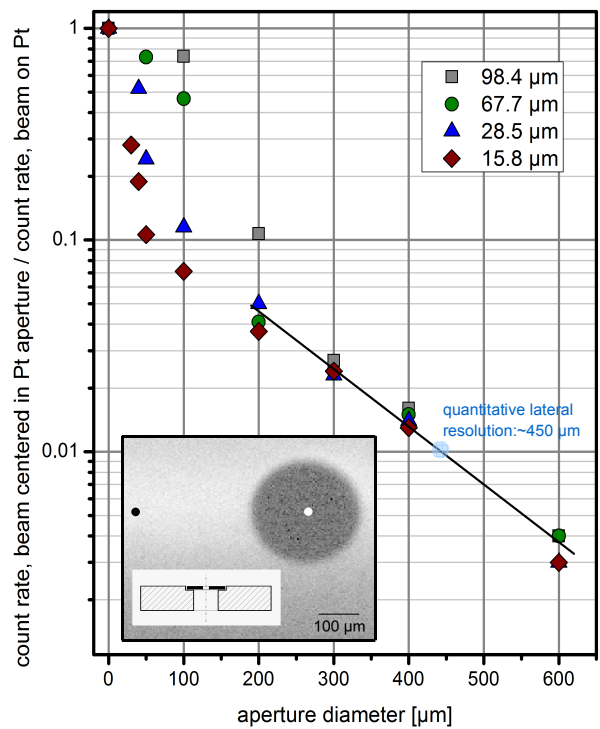

Fig. 3: count rate ratio as function of aperture diameter for X-ray beams with 4 different nominal beam diameters

The measured curves point out that two types of X-ray intensities contribute to the signal. The main contribution is given by the desired X-ray beam and additionally a slowly varying background is present. This slowly varying background is independent from the X-ray beam diameter. Most likely this background is due to the monochromator surface quality. Intensities near the direct scattering reflex are caused by of statistically distributed scattering centres. And the mutual distances of these centres have to be very large compared to the lattice parameter of the monochromator crystal.

The knowledge of the instruments quantitative lateral resolution is essential for the analysis of small sample features because it allows to estimate the intensity of small signals which are originated in the surrounding of the feature of interest.

## 4. Experimental Approaches

Due to the instrumental design the smallest area, which can be analyzed by an XPS microprobe with a focused X-ray beam is defined by the beams lateral dimensions. On one hand it is the nominal diameter of the X-ray beam. On the other hand there exists a trade off between a good lateral resolution and the detection of low elemental concentrations, due to the tailing of the X-ray beam intensity. As discussed above, the quantitative lateral resolution of an XPS microprobe can describe this property [1, 10].

Two experimental approaches are applied to examine sample structures independently, which are smaller than the X-ray beams dimensions. First, differential sample charging on partly non-conductive samples is utilized to discriminate conductive and non-conductive sample regions [2]. Second, by using the energy analyser acceptance volume method, only the sample is placed inside the energy analyser acceptance volume. This way, exclusively the sample contributes to the signal independent from the sample size if it is illuminated by X-rays.

## 5. Differential Sample Charging

The sample used to demonstrate the differential sample charging method is a microelectronic silicon device with Al bond pads of a size of approximately 120 x 120 $\mu m^2$. The silicon device was covered with a 30 $\mu m$ thick polyimide (PI) protective layer. The bond pads were opened by plasma etching of the PI. The quality of this etch process was controlled by XPS measurements using the Surface Science Laboratories X-probe instrument. The nominal X-ray beam diameter is 300 $\mu m$, which is quite larger than the bond pad.

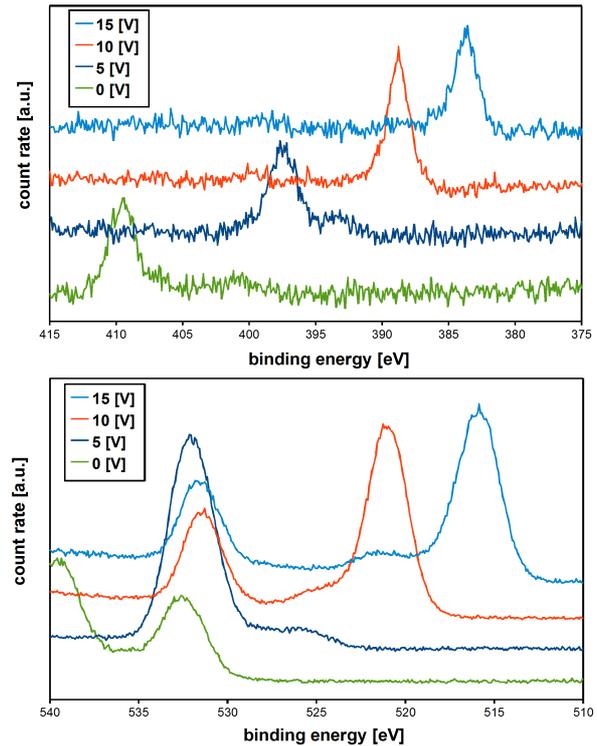

Fig. 4: high resolution N1s (top) and O1 (bottom) spectra, measured with the flood gun switched off or switched on using electron energies of 5, 10 and 15 eV

Fig. 4 shows the high resolution N1s and O1s spectra with the flood gun switched off or switched on using electron energies of 5, 10 and 15 eV, respectively. Both graphs show an overlay of spectra measured in different experiments. O is present in the PI and in the Al oxide at the bond pad surface. On the conductive bond pad surface the position of the O signal remains nearly constant in-dependent from the flood gun energy. Without energy flooding the PI becomes positively charged by photoelectron emission. The O1s signal of the PI is detected at a binding energy which is approximately 7.4 eV higher than the value measured on the Al bond pad. The non-conductive PI surface becomes negatively charged by the low energy electron flooding. The photoelectrons from PI are accelerated and therefore they are measured at lower binding energy. For a flood gun energy of 5 eV the O1s signals of the oxide and the PI superimpose. The N is only a component of the non-conductive PI because the peak position changes in all spectra. In the spectra some smaller peaks are visible indicating some limitations of the differential sample charging method. These additional peaks are attributed to particular surface areas. For example, the transition area between the conductive Al and the non-conductive PI at the edge of the bond pad may be a candidate.

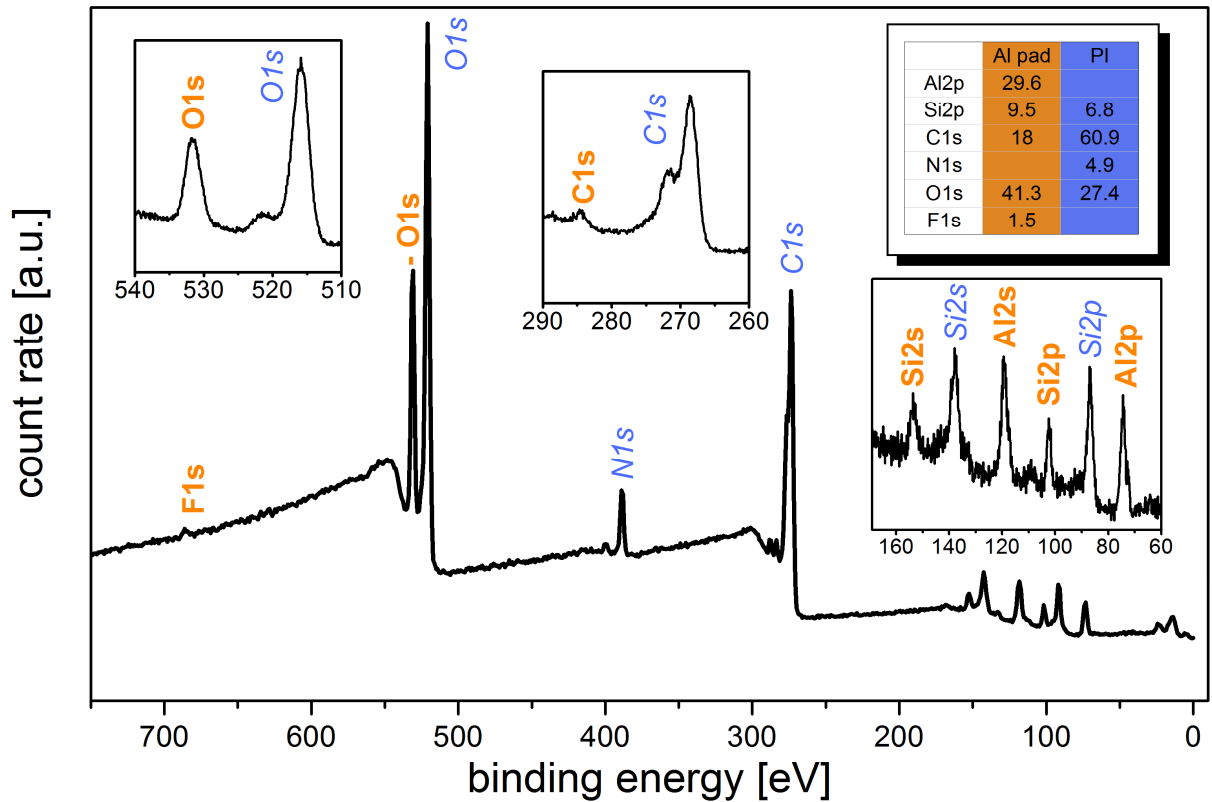

Fig. 5: survey scan of bond pad and PI measured with 15 V electron flood gun energy
Peaks belonging to the bond pad surface are labelled with normal/orange characters. Peaks of PI, which were labelled with italic/blue characters, are shifted to lower binding energy.

Fig. 5 depicts a survey scan of the bond pad and the PI recorded with 15 V flood gun energy. If a binding energy shift of approximately 15 eV towards lower binding energy is observed the measured intensities are assigned to the PI surface. The peak intensities attributed to the conductive Al bond pad appear at the expected binding energies. From the measured peak intensities the apparent atomic concentrations of the detected element for the PI and the Al bond pad were estimated separately. The calculation uses the instrumental sensitivity factors and assumes that the detected elements are distributed homogenously within the information volume, which is mainly defined by the information depth of the measurement. The table, which is inserted in fig. 5, summarises the results.

The Si on the PI is a hint to a residue of the mask which is necessary for the bond pad opening etching. As expected, the Al bond pad surface is oxidised. On the bond pad surface we detect a little amount of F, which is due to an etch residue of the bond pad opening etching. Additionally, we have some ambiguous C contamination on the bond pad surface. But there is no N on the bond pad detectable. This indicates that there are no PI residues at the bond pad surface and that the bond pad opening was perfect.

## 6. Energy Analyser Acceptance Volume Method

The second experimental approach uses the limited spatial extension of the energy analyser acceptance volume. Fig. 6 shows a plot of the analyser acceptance dependency of a Quantum 2000 as function of the z position of an Au foil. The z position gives the distance of the sample from the X-ray source since the incoming X-rays are parallel to the surface normal. The intensity of a normalised

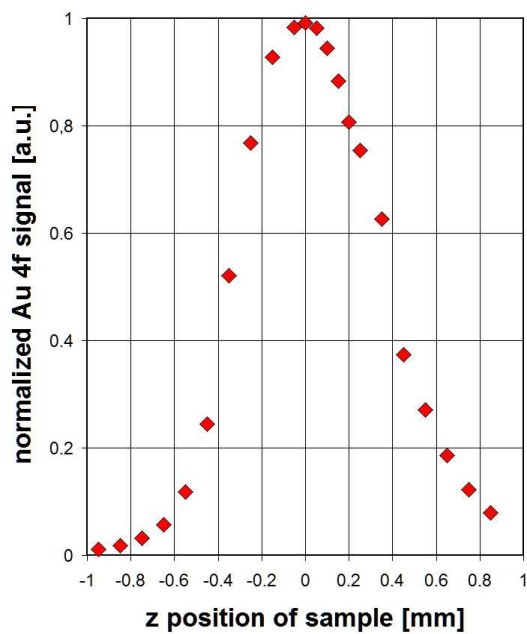

Fig. 6:   analyser acceptance dependency of a Quantum 2000 as function of the samples z position

Au 4f signal is plotted against the samples z position. Starting from the optimum sample position the intensity decreases drastically for a misalignment of less than 1 mm. Due to the rotationally symmetry of the analyser input lens the acceptance in x and y direction is expected to be comparable to this result. Therefore the analyser acceptance volume of the Quantum 2000 has an extent of a few cubic millimetres only. In praxis the analyser acceptance in the x-y-plane can not be determined by a measurement. This is unfeasible because electrostatic deflection plates at the energy analyser entrance synchronises the analyser acceptance with the position of the rastered X-ray beam on the sample surface by dynamic emittance matching [9].

The energy analyser acceptance volume method involves placing only the sample inside this volume. This way only the illuminated sample contributes to the measured photoelectrons intensity, independent of the sample size.

The analysis of Cu bond wire surfaces is discussed as an example of such a measurement approach. The bond wires have a diameter of approximately 80 μm. Fig. 7 shows the mounting of the bond wires. The bond wires are mounted completely free over a hole of the sample holder. The insert shows the experiment in detail. The measurements were done with a 100 μm spot size of the X-ray beam. From each wire, a survey spectrum of the surface 'as received' was recorded. From the detected peak intensities, the surface compositions of seven wires were estimated by standard data evaluation using the vendor's software package [11]. Tab. 1 summarizes the results. On all bond wires a higher amount of C was detected. Most likely, this C is due to the lubricant utilized during the wire drawing. Additionally, C contaminations are due to sample handling in ambient air. The detection of N at the surface of wire 3 is a strong hint to an antioxidant chemical on the surface of this wire [12]. On some wires, corrosive S and Cl contaminations are present.

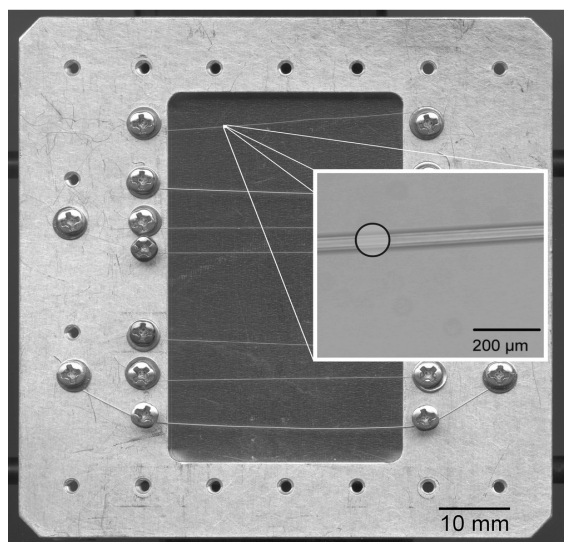

Fig. 7:   mounting of bond wires
The insert shows the dimensions of the bond wire relative to the 100μm X-ray beam utilized for analysis.

| sample | C1s | N1s | O1s | S2p | Cl2p | Cu2p3/2 |
|--------|------|-----|------|-----|------|---------|
| 1 | 42.9 | | 29.6 | | | 27.5 |
| 2 | 45.8 | | 34.7 | | | 19.5 |
| 3 | 42 | 6.7 | 33.1 | | | 18.2 |
| 4 | 52 | | 37.6 | | | 10.4 |
| 5 | 68.6 | | 25.2 | 1.7 | | 4.5 |
| 6 | 51.6 | | 35.8 | 2.3 | | 10.3 |
| 7 | 45.1 | | 31.6 | 2 | 1.1 | 20.2 |

Tab. 1:   Cu bond wire surfaces 'as received' elemental composition in apparent atomic concentration [at%]

Please notice, that the X-ray beam size is larger than the bond wire diameter. As demonstrated it was possible to perform a reliable measurement of the bond wires surface contaminations down to the 1% level, if only the sample and nothing else is placed within the energy analyser acceptance volume.

## 7. Summary

This article presents two experimental approaches, which overcome the lateral resolution and low elemental concentration limit defined by the X-ray beams lateral dimensions and tailing.

On partly non-conductive samples differential sample charging is used to separate spectra of different sample areas with dimensions below the X-ray beam diameter. Using low energy electrons with a kinetic energy in the range of a few electron volts non-conductive areas are charged negatively. The photoelectrons from this surface area are accelerated and therefore they are measured at lower binding energy.

The energy analyser acceptance volume method utilizes the rather limited dimensions of the energy analyser acceptance volume. Applying this approach only the sample and nothing else is mounted inside this volume. Therefore the sample has to have a suitable shape, which can be produced by a sample preparation if necessary. Important is the use of an adapted sample mounting where only the sample is probed.

## Acknowledgement

The measurements were done utilizing an XPS microprobe Quantum 2000 and a Surface Science Laboratories X-probe, respectively, installed at Siemens AG, Munich, Germany. I acknowledge the permission of the Siemens AG to use the measurement results here. For fruitful discussions and suggestions I would like to express my thanks to my colleagues. Special thanks also to Gabi for text editing.

---